%% file: ms.tex
\documentclass[twocolumn]{aastex631}

\usepackage{amsmath}
\usepackage{ulem}
\usepackage{xcolor}   
\usepackage{url}

\catcode`_=\active
\newcommand_[1]{\ensuremath{\sb{\mathrm{#1}}}}
\catcode`^=\active
\newcommand^[1]{\ensuremath{\sp{\mathrm{#1}}}}

\input defs

\shorttitle{}
\shortauthors{Judge et al.}

\newcommand{\hao}{
High Altitude Observatory,
National Center for Atmospheric Research,
Boulder CO 80307-3000,
 USA }

\begin{document}

\title{Optimal spectral lines for measuring chromospheric magnetic fields}


\correspondingauthor{P.  Judge}\affiliation{\hao}

\author{P.  Judge}\affiliation{\hao}
\author{P. Bryans}\affiliation{\hao}
\author{R. Casini}\affiliation{\hao}
\author{L. Kleint}\affiliation{University of Geneva, 7, route de Drize, 1227 Carouge, Switzerland}
\author{D. Lacatus}\affiliation{\hao}
\author{A. Paraschiv}\affiliation{\hao}
\author{D. Schmit}\affiliation{CIRES, 216 UCB, University of Colorado, Boulder, CO 80309, USA}

\date{Accepted . Received ; in original form }


%
%

\begin{abstract}
This paper identifies
spectral lines from X-ray to infrared wavelengths which are optimally suited to measuring vector 
magnetic fields as high as possible in the solar atmosphere. Instrumental and Earth's atmospheric 
properties, as well as solar abundances, atmospheric properties and elementary atomic physics are  considered 
without bias towards  particular wavelengths or diagnostic techniques.  
While narrowly-focused  investigations of 
 individual lines have been
reported in detail, no  assessment of the comparative merits of all lines has ever been published.  Although in the UV, on balance the Mg$^+$ $h$ and $k$ lines 
near 2800 \AA{} are optimally suited
to polarimetry of plasma near the base of the solar corona.  This
result was unanticipated, given that longer-wavelength lines offer greater sensitivity to the Zeeman
effect.  While these lines sample
optical depths photosphere
to the coronal base, we argue that
cores of \textit{multiple} spectral
lines provide a far more discriminating probe of magnetic structure 
as a function of optical depth than the core and inner wings of a strong line.
Thus, together with many chromospheric lines of Fe$^+$ between 2585 and the $h$ line at 2803 \AA{}, 
this UV region promises new 
discoveries concerning how the magnetic fields emerge, heat, and accelerate plasma as they battle to dominate the force and energy balance within the poorly-understood chromosphere.  
\end{abstract}


\section{Introduction}
\label{sec:statement}

The solar atmosphere, comprising those outermost regions from which 
radiation freely escapes into space, exhibits many phenomena related to
the emergence of magnetic fields generated in the  interior. While
less than 1 part in 1000 of the solar luminosity is associated with these  magnetic fields, their effects can be dramatic, being dynamically dominant  
in the tenuous chromospheric and coronal plasmas \cite[reviewed, for example, by ][]{Eddy2009}.
To understand mechanisms behind these phenomena, we 
must measure  magnetic fields above  the visible surface, the ``photosphere''.   We
are obliged to measure the magnetic
field \textit{vector}, not merely
components such as the vector projected on to the line-of-sight
(LOS), because the free magnetic energy driving flares and coronal mass ejections (CMEs) depends
on the electric currents  $\mathbf{j}=\nabla \times \mathbf{B}$ threading the plasmas.  

To determine  
$\mathbf{j}$ using spectral lines requires measurements of  linearly and circularly polarized line profiles, 
as well as unpolarized intensity  \citep{Landi+Landolfi2004}.   
While the line-of-sight component 
of $\textbf{j}$ can be fixed 
from measurements of $\textbf{B}$ 
in a plane perpendicular to the 
line-of-sight (LOS) \citep[e.g.][]{Pevtsov+Peregrud1990GMS....58..161P,Metcalf+others1994}, the perpendicular components 
require measuring features
 formed elsewhere along each LOS,  preferably in
a second spectral
line. Different
strategies can be used to map measured polarization states to 
magnetic field vectors. All require  
various assumptions and approximations. 
One strategy is to perform  ``inversions'', which iteratively modify a model magnetized atmosphere until a match to observations is found. In this way
\cite{Socas-Navarro2005a,Socas-Navarro2005b} inverted Zeeman-induced polarization observed in photospheric Fe and chromospheric Ca$^+$ 8542.1 \AA{} 
lines.  With his instrumentation and choice of
spectral lines, 
he could derive $\mathbf{j}$ only on  
scales $\ge$ 1Mm.  The measurement is crude because the entire stratified 
chromosphere spans only 1.5 Mm  \citep{Vernazza+Avrett+Loeser1981}.   \new{Examples} of more recent 
measurements of $\textbf{j}$ 
are photospheric work by \cite{PastorYabar+others2021} \new{and chromospheric work by 
\citet{Anan+others2021}.}

Common to all such
studies, the measurements must be  augmented 
with a variety of more-or-less 
credible assumptions and even ad-hoc procedures to ``fill in''  information missing from 
the data themselves.  ``Regularization'' is a well-known but subjective 
example employed in the inverse strategy \citep[e.g.][]{Craig+Brown1976}. Pure data therefore become laden with impurities, assumptions, some explicit (e.g., in the work of Socas-Navarro,
the atmosphere is in
hydrostatic equilibrium), others implicit (e.g., use of a small number of ``nodes'' at which numerical solutions for physical quantities are sought). However, the 
chromosphere 
is incompatible with many  
simplifying 
assumptions 
commonly applied to
the photosphere (LTE, Milne-Eddington 
solutions, statistical equilibrium) 
or the corona (optically thin).  Our understanding  of chromospheric magnetic fields 
and their interaction with plasma 
is accordingly 
rudimentary.  We
do not yet know how the chromosphere modulates magnetic 
energy emerging through the photosphere before reaching the corona.

Given these challenges,  we undertook a search for information-rich  lines in the solar spectrum, in the hope of
maximizing the 
magnetic information content of
remotely sensed 
magnetic fields. 
Our search is 
conducted  without 
bias towards specific wavelengths or observing platforms, \new{ for the first time.}

\section{Requirements}
\label{sec:requirements}

We seek spectral lines only of atoms and atomic ions.   Molecules, useful in the deep photosphere, will be ignored
because their population densities
drop to negligible values, exponentially with half of the pressure scale height 
of the plasma (itself just $\approx 150$ km).  
Photospheric magnetic fields vary from a few G, 
to as much as 
several thousand G in
the cores of sunspots. Being
farther from sub-photospheric sources of electric currents, chromospheric magnetic fields generally are weaker.

The lines
of most interest must:
\begin{itemize}
    \item  be able to reveal signatures of solar magnetic fields down to 1G, without specialization to a specific physical origin,
    \item be able to constrain the vector field \textbf{B}, not just a component of it,
     \item be sufficiently opaque to form within the higher, more tenuous plasma in and above the solar chromosphere,
    \item have opacities simply related to atmospheric pressures and temperatures, to fix relative heights of formation, 
    \item be readily observable with current instruments,
      \item be bright enough to achieve signal-to-noise ratios sufficient to interpret the magnetic signatures.
\end{itemize}

The final criterion is of first importance, because spectropolarimetry is photon-starved, even on the Sun's bright disk  
\citep{Landi2013}.
The third criterion
demands the use of very strong (i.e. opaque) lines, which 
traditionally have 
been poorly
understood.  However,
recent work has shed light on 
the formation of both intensity 
\citep{Judge+others2020} and polarization
\citep{Manso-Sainz+others2019} of such lines. 
The fourth criterion is not satisfied for 
lines of H and He between any excited levels (e.g. H Balmer 
$\alpha$ and He 10830 \AA). 

Generally, any measured signals  should also be stable against 
instrumental fluctuations and 
degradation, and the negative effects of a variable terrestrial atmosphere (extinction, scattering, 
seeing).   Polarimetry requires multiplexed
simultaneous measurements 
to avoid spurious noise and 
cross-talk, setting requirements on integration times for modulation and demodulation,  
which can conflict with 
rapid evolution of the solar
plasmas and SNR requirements
\citep{Lites1987,Judge+others2004,Casini+others2012}.
Continuity and regular repeatability of observations
over periods up to months is 
highly desirable, to follow evolution of emerging fields, 
wave motions, and the explosive release of magnetic free energy stored within the atmosphere 
over extended periods. The evolution of
magnetism in active regions and coronal holes, for 
example, demands extended observations over hours to months.   The day/night and seeing cycles, as well as weather events,
present challenges at
most ground-based observatories. Thus 
operations in space, necessary for
UV and some infrared observations, should perhaps be 
favored. 

These considerations  point to the primary requirement for this study:
\textit{the spectral lines must be as bright as possible.}

\section{Solar and atomic properties}

\subsection{Elemental abundances}

The need for strong, opaque  spectral
lines limits the selection
to a subset of roughly 
10-15 elements. Those with  abundances above $10^{-5}$ of hydrogen are  \citep{Allen1973}:
\begin{verbatim}
H, He, 
C, N, O, Ne, 
Mg, Si, S, Ar,
Ca, Fe,
\end{verbatim}
where the elements are grouped according to the row that they occupy in the periodic table.
Of these, all elements 
have stable dominant isotopes with even numbers 
of nucleons except for 
H and N. Consequently, 
both H and N have nuclear-spin-induced hyperfine structure, introducing higher  levels of complexity 
into current models, which have yet to be generally implemented 
\citep{Alsina+others2016}.

\figstart

\subsection{Atmosphere}

The Sun's atmosphere is a continuously evolving, partially or fully ionized plasma threaded by magnetic fields.   The pressure
in the line-forming region of the photosphere happens to be about $10^4$ dyne~cm$^{-2}$, in the corona it varies from 
about $0.2$ 
to  10 dyne~cm$^{-2}$ under conditions of 
extreme heating in active regions.  Temperatures in
these regions are $\approx$5000 K (upper photosphere), and 
from $10^6$ to say $3\times10^6$ K (corona), outside of  flares.  

The intervening 
chromosphere, spanning 
about 9 pressure scale heights, 
is stable to temperature fluctuations driven by over- heating, most excess energy going to latent heat of ionization, powerful radiation losses and some into 
plasma motions.  Standard models \citep[e.g.][]{Vernazza+Avrett+Loeser1981} place its temperature near 6000-8000
K ($\equiv 0.5$ eV) across these changes in
pressure, with an almost
constant electron pressure of $0.1$ dyne~cm$^{-2}$ but with partial pressures of 
neutral hydrogen
varying from $1000$
to $0.1$ dyne~cm$^{-3}$.

The average pressure drop
with height
has a
direct impact on 
our selection of spectral lines. The 
vast bulk of the chromosphere must be close to hydrostatically
stratified given the 
observed sub-sonic motions (outside of
a tiny fraction of area from which  
spicules emerge, \citealp{Judge2010}). Thus, on average, 
pressure
supports the weight of material above it:
\begin{equation}
    p = m g
\end{equation}
where $m=\int \rho dz$ (g~cm$^{-2}$) is the column mass, and $g$ the solar
acceleration due to gravity.   
The line center optical
depths $\tau_0$ of many spectral
lines are indeed simply related to
the column mass $m$ (fourth 
requirement listed in section
\protect\ref{sec:requirements}).
For a given internal
state of excitation of
the radiating or absorbing ion
\citep[][ignoring stimulated emission, and using standard notation]{Mihalas1978}:  
\begin{eqnarray}
    \tau_0 
    &=& \frac{\pi e^2}{m_ec} f_{ij} \
    \int n_i \phi(\nu_0) dz,\  \mathrm{where} \label{eq:atmos1}
    \\
        \int n_i dz 
    &=& \int 
\left\{
    \frac{n_i}{n_{ion}}
    \frac{n_{ion}}{n_{el}}
    \frac{n_{el}}{n_H} 
    \frac{n_H}{\rho} \right \}\ \rho dz, \label{eq:atmos2}
\end{eqnarray}
where $\phi(\nu_0)$ is the 
line profile function at line center frequency $\nu_0$, and 
the term in braces simply expresses the number density in the 
lower level as a fraction of the total
plasma density. If both quantities are roughly constant along
the line of sight $z$, then $\tau_0 \propto m$, thereby satisfying 
requirement 4 of
section \protect\ref{sec:requirements}. Not all lines in the solar
spectrum satisfy this criterion.

Between
chromospheric and coronal plasma is the intermediate temperature ``transition region''  with very little plasma owing to the large 
average temperature gradient there\footnote{Two kinds of structure are known to contribute to the transition region emission from ions such as He~II, C~IV, O~VI evident in Figure~\ref{fig:start}, depending on
whether the plasma is energetically connected to the corona. In both cases little plasma is present  \citep{Judge2021}.}.  The majority of ``transition region''
lines are optically thin, with the exceptions 
of resonance transitions of
H and He.  

\subsection{Spectrum}

Most
spectral lines observed from visible to infrared
wavelengths (henceforth ``vis-IR") form as narrow absorption features in the photosphere.    Near-LTE conditions characteristic of high pressure plasma 
ensure that the dominant 
ionization stages are neutrals (H, He, O, Ne) or, for elements with lower ionization potentials, singly ionized 
(e.g., Mg$^+$, Ca$^+$, Fe$^+$).
Between
the lines 
is the near-black-body continuum,  formed 
as the H$^-$ ion becomes 
optically thin at plasma
pressures near 10$^5$ dyne~cm$^{-2}$, with 
a characteristic radiation
temperature near 5770 K.

Vis-IR lines are generally weak, because
they mostly originate from between excited atomic  levels, with small populations and  little opacity.  
The line spectrum of neutral iron 
dominates by number the 
visual solar spectrum, yet even the resonance lines 
($4s-4p$ transition array)
near 3720 \AA, are a factor of 
500 less opaque
than the $4s-4p$ 
transitions in Fe$^+$ near 2600 \AA, because of smaller
oscillator strengths (see below) and different chromospheric 
ion populations $n_{ion}/n_{Fe}$.

One exception is hydrogen, which is abundant enough to generate significant photospheric opacity in transitions between excited levels (with principal quantum number $n \ge 2$) to generate deep absorption lines.  Resonance lines
lie at UV and vacuum UV wavelengths, except 
in lithium, sodium and potassium-like ions.  Their $ns-np$ transitions (note the same value of $n$)  are strong
electric dipole (E1) transitions between levels
separated only by the residual electrostatic interaction, and not the central Coulomb field. The resonance lines of  Li-like sequence of ions are clearly 
identified as C~IV, N~V, O~VI, Ne~VIII and Mg~X in Figure~\ref{fig:start}.
The $3s-3p$ sodium D lines lie at 5800 \AA,
and the Na-like Ca$^+$ $H$ and $K$ lines, the two strongest lines in the vis-UV spectrum, lie at 3969 and 3934 \AA.  (Throughout we use air wavelengths above 2000 \AA, vacuum below). 
Ca$^+$ also
has a trio of less opaque lines
near 8542 \AA, sharing 
a common upper level with $H$ and $K$, 
decaying to a metastable level 
1.7eV above the 
ground levels. 
These lines have
cores forming 
in the mid chromosphere
($p \approx $ 1 dyne~cm$^{-2}$). 

At wavelengths below the
Balmer continuum (3650 \AA), 
absorption lines of complex
ions of the iron group 
dominate in number, and they generally have  higher opacities than  vis-IR
transitions. 
The unfilled $3d$ shell has many meta-stable
levels as the equivalent  
hydrogen's 
``ground level'' is split 
into multiple levels with the same parity.  These levels are highly  populated, with modest energies $\approx  1-2$ eV above  the 
lowest level.

Below about 2000 \AA, the spectrum changes character as line opacities become 
mixed with 
photoionization continuum opacities of abundant elements, formed well above the photosphere near $p\approx 10^{3}$ dyne~cm$^{-2}$ \citep{Vernazza+Avrett+Loeser1981}.  The spectrum 
is superposed
with the well-known emission lines
originating from heated and warmer plasma lying above the continuum formation regions, and at yet lower pressures.  

Table~\ref{tab:calcs} lists only the brightest lines in the solar spectrum which are formed well above the photosphere, 
for further scrutiny below. 
Absent are well-known weaker 
magnetically-sensitive 
lines of abundant neutrals of the iron group, and also even  ``chromospheric''
resonance lines of neutrals, such as Ca 4227 \AA{}. Owing largely
to ionization balance (see below), the latter is formed in dense, high pressure 
plasma ($p_{plasma}\approx 20$ dyne~cm$^{-2}$), typically 1000 km beneath  
the corona
\citep{Bianda+others2011}.

\subsection{States of ionization}

The requirement to use the 
strongest lines has consequences dictated by the solar 
plasma temperatures, and densities, which primarily determine the dominant stages of ionization in the atmosphere
(factor $n_{ion}/n_{el}$ in equation
\ref{eq:atmos2}).  In the 
deepest, highest pressure chromospheric layers, 
ionization equilibrium is almost 
controlled through the principle of
micro-reversibility leading to LTE. The ionization occurs 
via impact of ambient
plasma electrons on atomic ions (one electron before, two after the collision), balanced by the 
reverse thermodynamic process of 3-body recombination (two electrons impacting the ion, one free electron after the collision).
The ionization follows
Saha's formula, and 
consequently neutral atoms with 
ionization potentials 
less than about 8eV tend to
be fully ionized. 
For other abundant elements, neutral species are also ionized by  solar radiation 
at UV and EUV wavelengths.
Except for H, He and some noble
gases with high ionization potentials, the dominant stage of ionization across the 
chromosphere is X$^+$ 
for most elements X. 

The spectral lines of interest form in higher, much less dense plasmas where LTE   
does not apply. Only 2-body collisions are then 
important, so that ionization by electron impact is balanced by radiative plus dielectronic recombination. 
In this ``coronal'' regime, ionization fractions 
 are essentially 
simple functions of electron temperature
\citep{Woolley+Allen1948}). 
The LTE and coronal ionization regimes are
clearly illustrated 
in an early review by  \cite{Cooper1966}.
As a rule of thumb, 
atomic ions with net 
charge $z$  have a maximum abundance near electron temperatures
\begin{equation} \label{eq:ioneq}
    T_e \approx 10^4 z^2, 
\end{equation}
a rough approximation  when 
$T_e > 10^4$ K, 
based upon general formulae
balancing electron impact ionization and radiative recombination.  Thus, 3$\times$ ionized 
ions would form near $10^5$K, those $10\times$ ionized near $10^6$K, temperatures of the mid transition region and quiet corona respectively.  

\subsection{Atomic physics}

For solar applications we seek
radiative signatures from 
 fields as weak as $1$ G,
and as strong as a few kG,
within the chromosphere and
corona.   A  field of magnitude $B = |\mathbf{B}|$ 
acting on a (non-degenerate) level produces an energy splitting 
of order $\mu_B B$ (Zeeman splitting) where
$\mu_B$ is the Bohr magneton, and 
an alignment of the gyrating atom along $\mathbf{B}$. In the absence
of spin-induced 
hyperfine structure (elements H, N)
or anomalously close levels, the fine-structure splitting greatly exceeds
the Zeeman splitting for most solar
magnetic fields. In such cases,
one can exploit
the familiar perturbation
formulation of the Zeeman effect. \cite{Cowan1981} lists  
dependencies of fine-structure splittings under
various scenarios, also as 
a function of net charge $z$, varying as $(z+1)^\alpha$ with various values of $\alpha > 2$.   

The magnetic splitting and associated alignment (breaking of symmetry) leads to 
polarized spectral lines even if the lines are far narrower than the thermal line width.   Then, the magnitude of the Zeeman-induced polarization depends on the quantity $\varepsilon\approx
    \mu_B B /\Delta E$, where $\Delta E$ is the 
width of the spectral
line in energy units.  For 
a plasma temperature of $T$ K and an ion of mass $m$, we have for a Doppler dominated line  
\begin{equation}
    \varepsilon \approx
    \frac{\mu_B B }{\Delta E} 
    =         \frac{\mu_B B }{\sqrt{kT/m}}\frac{\lambda}{h}  \ll 1.
\label{eq:zeeman}
\end{equation} For energy splittings small compared with fine structure and line widths, Zeeman-induced circularly
polarized light varies as $\varepsilon^1$, linear polarization varies as $\varepsilon^2$ \citep{Landi+Landolfi2004}.  
This familiar result favors 
longer-wavelength lines of heavier ions formed in lower temperature plasmas. All three factors lead to  a higher sensitivity to the Zeeman effect.  

Sometimes $\mu_B B$  is
comparable to fine structure energies when accidental 
atomic level crossings 
are present, as occurs in some complex ions. 
In this case,  the mixing of two otherwise 
``pure'' atomic states by 
even a modest magnetic field 
can give rise to an otherwise completely absent spectral line (magnetically induced transition, or MIT).
The new line's intensity 
varies as $B^2$. 
Such a mixing has led to a proposal  
\citep{Grumer+others2014,Li+others2016,Si+others2020}
to examine the EUV spectrum of chlorine-like ion Fe~X, and
\cite{Si+others2020a,Landi+others2020} have recently used this method to 
estimate magnetic field strengths in coronal
plasmas.  
This method has several practical challenges such as
line blending in the EUV, sensitivity to  $B=|{\mathbf B}|$ and using lines which are
optically thin so that information along the LOS is 
almost absent. Thus
we seek other diagnostic techniques.

\figcomp

The less familiar Hanle effect exploits 
the modification by magnetic fields of \textit{pre-existing} spectral-line polarization produced by resonance scattering. Scattering 
polarization originates even
in the absence of external
fields, through excitation processes that are
anisotropic, such as irradiation of atoms by light coming from a preferred direction, or by 
non-isotropic collisions.  Further, the associated atomic sub-levels evolve coherently as their wavefunctions are mixed
(entangled), their 
levels being degenerate 
in the absence of external fields. 

While a rigorous treatment of the Hanle effect requires quantum electrodynamics \citep{Landi+Landolfi2004}, a classical picture illustrates the energy regime in which magnetic fields  influence the polarization of the scattered radiation. During a time of order $A^{-1}$ necessary for an excited level 
to decay radiatively (where $A$ is the Einstein coefficient for spontaneous emission), the classical damped electron oscillator precesses around the magnetic field with the Larmor frequency \begin{equation}
\omega_B=\frac{\mu_B}{\hbar}B,
\end{equation}
 where
$\mu_B$ is Bohr's magneton,  Then the mean linear polarization of the radiation emitted by the damped oscillator will generally be rotated and reduced in amplitude with respect to the field-free case. The Hanle effect
is thus sensitive to magnetic field strengths $B$ of order 
\begin{equation}
    B_H \sim \frac{\hbar A}{\mu_B}
\label{eq:hanle}
\end{equation}
With $A\sim10^{8}$ s$^{-1}$ for
a typical E1 transition, $\mu_B \approx 10^{-20}$ erg~G$^{-1}$, then %
\begin{equation}
B_H \sim 10^{-7} A\, \ \ \ \mathrm{Gauss}.
\end{equation}
For lines of ions in an isoelectronic sequence, $B_H$ 
values can be estimated 
in plasmas at temperatures
given by eq.~(\ref{eq:ioneq})
from 
\begin{equation}
A(z) \approx A(z=0)z^\alpha,
\end{equation}
with $\alpha=4$ or 2 for 
E1 transitions with (a or no)  change in 
principal quantum number
respectively.  In the L$_\alpha$ line of He$^{+}$, formed near 10$^5$ K for
example, the critical fields are larger than for H L$_\alpha$ by a factor of 16
(see Table~\ref{tab:calcs}, which also contains
$\overline G$, the Zeeman
factor for linear polarization
of \citealp{Landi+Landolfi2004})\footnote{The classical argument permits an 
understanding of the Hanle effect
 in terms of 
polarization changes arising during the decay of the upper level only.  There is also a lower level effect where
the polarized scattered light 
is modified during the excitation 
of the gyrating atom from the lower level. Instead of the A-coefficient, the critical field strength depends on the Einstein $B-$coefficient and the incident radiation intensity, such that $A$ is replaced by $BJ$ 
where $J$ is the mean incident intensity.  Usually $A \gg BJ$ in the Sun, except perhaps at infrared wavelengths.  }.

\section{Analysis}

In bringing together 
the previous sections we can identify the most promising spectral lines for measuring those magnetic configurations well above the photosphere, which cause  flares, heating, and plasma eruptions.  

\subsection{A first cut}
 Figure~\ref{fig:comp} shows   estimates of the line core intensities, in photon units, compiled 
in Table~\ref{tab:calcs}. The cores are our focus for two reasons: they form highest, and the Hanle effect operates in the Doppler cores.
By far the  lines with highest photon fluxes
are, in order, 
He 10830, 
Mg$^+$ 2795 and  2803,
H B$\alpha$ 6563,
Ca$^+$ 3933 and 8542, 
H L$\alpha$ 1216, a multiplet of Si$+$ whose strongest 
line is at 1817 \AA, and resonance 
lines of Fe$^+$ between 2585 and 2632  \AA. 
These are all
\textit{chromospheric} lines, not surprising because the chromosphere
radiates far more energy than the transition region and corona
\citep{Withbroe+Noyes1977}.  Our first
cut therefore includes only these lines, all
of which, when observed against the solar disk, are optically thick in the chromosphere. The exception is He 10830, a
weak absorption line, barely visible outside
of active regions, whose 
origin is still a subject
for debate \nnew{(compare abstracts of papers by \citealp{Judge+Pietarila2004} 
``the spatio-temporal properties of the He I 584 \AA\ldots{} are qualitatively unlike other chromospheric and transition region lines'',
and that of
\citealp{Leenaarts+others2016}), ``the basic formation of the line in one-dimensional models is well understood''.}
Although the first  paper focuses on resonance lines and 
the latter on 10830, both EUV and IR lines of helium require excitation to 
levels
inaccessible to electrons at local thermal energies. 
\nnew{Further, \citet{Leenaarts+others2016} did not recognize  
the failure of one-dimensional
models to explain anomalous Doppler shifts of the 10830 \AA{} line
\citep{Fleck+others1994}
which are in phase with 
the well understood 
\ion{Ca}{2} $K$ line. }
The optical depths listed in Table~\ref{tab:calcs} can be scaled to solar values, they are all relative to H L$\alpha$, computed from identical geometrical depth of 100 km and 
electron and hydrogen densities of 10$^{11}$ particles cm$^{-3}$ with the additional
assumption that each ion listed 
has $n_{ion}/n_{el}=1$
(equation~\ref{eq:atmos2}). 
No microturbulence
was included in the line profiles.   In the Sun
the plasma path lengths vary from Mm
in the chromosphere, which also has higher neutral densities, 
to below 100 km in the steepest 
part of the transition region, with at least 10$\times$ smaller densities.   The arbitrary choice 
to normalize optical depths 
using a path length of 100 km is
useful, but one must remember to
scale optical depths from those given in the table.

Of the strong lines, we discount 
that of Si$^+$ owing to an unusually low oscillator strength, consequently an optical depth 500 times smaller than Mg$^+$, and 
a low Hanle field strength $B_H$.

\tabstrong

\subsection{H Lyman and Balmer series}

H L$\alpha$ has been fruitfully used to 
determine plasma properties including magnetic
fields using the Hanle effect, relatively high in coronal plasma \citep[e.g., see the review by ][]{Raouafi+others2016}, and more recently in the chromosphere, using data from the rocket experiment of the Chromosphere Lyman-Alpha Spectro-Polarimeter \nnew{(CLASP, \citealp{Ishikawa+others2011,Trujillo2018})}.  The Zeeman effect is ill-suited owing to the short wavelength,
moderate photon flux, and thermally broad
lines.   Below 2000 \AA{}, instrument 
degradation by 
stray contaminants exposed to intense solar radiation 
is \new{a concern, but available data are difficult to
assess. The IRIS UV channel centered at 
1335 \AA{} degraded by a factor of 8 over 4 years, compared with a factor 1.3
at 2800 \AA{} \citep[see Figure 24 of][]{Wuelser+others2018}. This degradation occurred predominantly in the first months apparently due incomplete outgassing and is related more to detector issues than optics  \citep{Wuelser+others2018}. 
The SUMER instrument on SOHO (in orbit around L1) also experienced degradadation, although in a lesser degree. The  overall degradation was about 20\% over 3 years, most likely due to  contaminants on optics \citep{lemaire2002}. The origin of these degradations might also be related to the SOHO attitude loss event. Thus, far UV degradation might be mitigated even at shorter UV wavelenghts.}
There are few 
data on polarization characteristics  of such issues 
\citep{Santi+others2021}.  \new{\citet{Alsina+others2022} recently made 
calculations of the Lyman $\alpha$ wings
including magneto-optical effects.  They showed that linear polarization of a few percent in $P/I$ in the wings ($\pm 5 $\AA) is
sensitive to magnetic fields of order $100$ G or higher.  The wings form in the middle 
chromosphere \citep{Vernazza+Avrett+Loeser1981}, thus
the line should be useful to probe the stronger magnetic fields in active regions.
}

The \new{six} blended components of H Balmer $\alpha$ are strong, with a combined line center depth 0.17 of the continuum.  Line wings form in the photosphere, the core forms in the upper chromosphere.  They pose 
significant challenges for
Zeeman and Hanle polarimetry, in addition to
blending. Firstly,  the 
optical depth scale  
depends not on column mass
$m \propto n_H(n=1)+n_p$ (equation~\ref{eq:atmos2},  
population densities of the $n=1$ level of
neutral H and protons respectively), but on $n_H(n=2)$ levels lying 10.2 eV above the $n=1$ level.  The $n=2$ level
populations are sensitive to
non-LTE and also non-equilibrium effects, owing to hydrogen's  unusually small rate of recombination \citep[explained in basic terms by][]{Judge2005}. Within the chromosphere it takes $\approx 10^2$ seconds for 
hydrogen to recombine once ionized (the Lyman continuum ionization and recombination rates cancel,    ``case B''  of \citealp{Baker+Menzel1938}),  comparable to
the natural chromospheric oscillation period of 3 minutes. 
The degeneracy of the levels with respect to angular momentum quantum numbers leads to more,  well-known complications, such as 
a linear Stark effect which broadens the lines in plasmas even above the already thermally broadened lines. Both the Zeeman and Hanle effects  found little if any success in measuring magnetic fields using any Balmer lines,
\new{until recently.
\citet{JaumeBestard+others2022} used the 
unique ZIMPOL 
instrument to 
record Stokes signals of
H$\alpha$ at a very low noise level ($10^{-4}$ of continuum brightness) in 5-8 minutes with a 45 cm
telescope.  Fractional Stokes polarization  signals of $\approx$0.1\%{} were
achieved by binning over 8 detector pixels.  All in all, while interesting, the currently low signals that hinder interpretation together with the 
poorly-constrained  formation conditions of H$\alpha$ make it less desirable as a prime diagnostic of vector magnetic fields.}

\subsection{He 10830}

The helium line at 10830 \AA\ has been 
usefully employed by many authors to diagnose solar plasma including magnetic fields. Codes have even been developed to invert polarized light 
from this transition, albeit with highly simplified  assumptions 
\citep[][who adopt a slab geometry]{Hazel}, \citep[][a simplifying Milne-Eddington approach]{2004A&A...414.1109L}. These approaches yield essentially
no information on the heights of formation of the
lines, except high above the photosphere far from our regions of interest.
Like H Balmer lines, the $1s2s ^3S$
and 
$1s2p ^3P$ lower levels lie 20 eV above the singlet ground state, rendering optical depths again sensitive to nLTE, non-equilibrium effects.  Further,
the 20 eV energy of the lower level contrasts with  the mean kinetic energy of
electrons below $2\times10^4$ K, rendering the lower level populations and hence opacities
sensitive to high energy tails of  distributions  in electrons and photons, and dynamical non-equilibrium effects, again 
like hydrogen
\citep{Pietarila+Judge2004}.  

In short, the heights of formation of the 10830 line are 
essentially undetermined.     Also like hydrogen,
the low atomic mass generates broad lines, 
lowering Zeeman polarization signals (equation~\ref{eq:zeeman}). 
The critical Hanle field strength is just 0.87G, making this line insensitive to magnetic fields over active regions where the line
is strongest, and perhaps
most interesting
for energetic events. 

\subsection{Mg II vs. Ca II}

We are left to discuss the strong lines of
Mg$^+$ and Ca$^+$. The term
structure of these two elements is similar but
with notable differences.
Ca$^+$ is K-like 
with a $n=4$ outer shell, so 
that
between the $3p^64s~^2S_{1/2}$ ground level and $3p^64p~^2P^o_{1/2,3/2}$ upper levels of the $H$ and $K$ lines, there are 
two metastable levels. These 
$3p^63d ^2D_{1/2,3/2}$ levels lie near 1.7eV above the  ground
level,  no equivalent levels can exist in Mg$^+$ which has no $2p^6 2d$ sub-shell.  Consequently, the lower
level populations of Ca$^+$ 
transitions are spread
among these 3 levels, and the individual line opacities are lower per ion than in Mg$^+$.  The 
8542 \AA{} line ($3p^64p~^2P^o_{3/2}$
 to 
$3p^63d~^2D_{5/2}$) has been used frequently for chromospheric 
spectropolarimetry, but 
the 1.7 eV lower level energies lead to opacities
smaller than $H$ and $K$ by a factor of 20.  Mg$^+$ ions
also survive  in plasmas with higher electron temperatures with higher EUV radiation levels than Ca$^+$, owing to the 3.1 eV
difference in ionization potentials.  Ca$^+$ is 50\%{} ionized at about 8500 K, Mg$^+$ at 17,000 K in the models of \cite{Vernazza+Avrett+Loeser1981}.

Other considerations might favor
the Ca$^+$ transitions:
Mg$^+$ observations must be conducted from a very high
altitude balloon \citep{Staath+Lemaire1995} or from space, whereas 
all the Ca$^+$ lines are observed from the ground.
Ground-based observations of the $H$ and $K$ lines 
(3968\,\AA\ and 3934\,\AA, respectively)
are notoriously difficult to correct for atmospheric seeing compared with those obtained at longer 
wavelengths. This can reduce  the spatial resolution advantage offered by shorter visible wavelengths at the $H$ and $K$ lines. Figure~11 of
the instrument paper for the Visible Spectro-Polarimeter on DKIST 
\citep{deWijn+others2022}
clearly illustrates 
this problem:  the simultaneous scans of the photospheric continuum around the three observed wavelengths of Ca~II H, Fe~I 6302, and Ca~II 8542, show the stronger influence of seeing, plus the lower efficiency of adaptive optics corrections at shorter wavelengths.  DKIST adaptive optics are optimized for wavelengths around 4750-5750 \AA\ \citep{rimmeleaobook} and seeing (or more exactly the Fried parameter) is proportional to $\lambda^{6/5}$ \citep{rimmele2011}.


The relatively 
low photon fluxes in these line cores (Figure~\ref{fig:comp})
demand larger telescope apertures where seeing effects become more challenging for fixed seeing conditions, and/or longer 
 integration times.  Both
 of these demands conflict with the fast modulation cycles needed to reduce 
 noise and crosstalk inherent in polarization
 measurements which must be made before the solar 
 features themselves also change \citep[e.g.][]{Judge+others2014}.  These arguments would seem to dictate use of a space platform for the $H$ and $K$ lines, even though the strong Ca$^+$ chromospheric lines are visible from the ground.   

In conclusion, if Ca$^+$ requires a space platform, then one may
as well immediately turn to Mg$^+$, at least an order of magnitude
more opaque and with opacity extending into plasma hotter by 
a factor of 2.  Further, lines of Fe$^+$ exist with diverse opacities similar to and less than those of Ca$^+$, but at wavelengths relatively close to Mg$^+$.
While  \cite{Judge+others2021} argued for the combination of Mg$^+$ and Fe$^+$ lines at wavelengths between 2560 and 
2810 \AA{} as a
critical wavelength range to perform chromospheric polarimetry, here we confirm and extend this result to show that it is also \textit{optimal}, no matter the wavelength region selected for observations.

While it is clear that Mg$^+$ $k$ is the optimal single line, and that a 2-level treatment 
of the radiative scattering problem in this ion might 
be applied to the Hanle effect, the 
Ca$^+$ and Fe$^+$ ions 
require a multi-level treatment.  The scale of calculations needed to supplement Mg$^+$ work  increases 
far more than quadratically
than the square of the number of levels (such as 5 for 
Ca$^+$, many more for Fe$^+$), owing to the 
need to solve for off-diagonal elements of the atomic density matrices
in statistical equilibrium
\citep{Landi+Landolfi2004}.
Such large-scale calculations are not currently 
routine, but are being implemented
\citep[e.g.][]{Li+others2022}.

\section{conclusions}

The best lines for chromospheric polarimetry 
 in the whole X-ray-infrared solar
spectrum are the resonance lines
of Mg$^+$.   Quantitative polarimetry 
using these lines was already 
achieved in the 1980s with the UVSP
instrument on SMM \citep{Henze+Stenflo1987}. It has received yet more attention with 
a recent CLASP-2 rocket flight \new{
\citep{Ishikawa+others2021,Rachmeler+others2022}. Data from this flight
demonstrated the first qualitative 
agreement of existing 
theories of 
scattering polarization
in these strong lines, 
 \citep{Belluzzi+TrujilloBueno2012}, 
modifications to $k$ line core polarization
due to the Hanle effect
\citep{Alsina+others2016,delPinoAleman:2016,delPinoAleman:2020},}
and emphasized the 
suitability
of these lines \new{for polarimetry using }  modest
telescopes, and just 150 seconds of
exposure time.

\new{However, the interpretation 
of Hanle signals
from the core of one strong line presents known difficulties.   
In analyzing H Lyman $\alpha$ data from
the first CLASP 
flight, 
\citet{Trujillo2018}
demonstrated the failure of 
1D models to account for center-to-limb 
behavior of the 
polarized line core.
Thus they were obliged to 
propose non-planar
corrugated optical surfaces to explain the observations.  
Such hypotheses 
can be explored further by using
multiple lines.} While 
a single line's profile does form over  multiple depths, the source functions of  strong  lines are  non-locally controlled
\citep{Mihalas1978,Ayres1979}, being influenced by radiation from a huge
range in optical depth. \new{
Ideally a series of lines with different opacities is desirable to constrain and better understand the 
hydromagnetic state across 
the chromosphere.   This approach was suggested 
with particular use of
the Mg$^+$ lines, shown here to
be optimal, together with neighboring lines of 
Fe$^+$, all lying between 2585
and 2810 \AA{}, by \citet{Judge+others2021}}

\new{ Observed together with 
these other}
lines,  the Mg$^+$ lines
have compelling advantages:
large photon fluxes, 
reliable knowledge of relative depths of formation, 
the minor role of blends, combined sensitivity to
Hanle and Zeeman effects, and continuity of time-series data
\new{that accompanies stable space platforms.} Such observations  hold potential
to answer many outstanding 
questions arising from 
several ground- and space- based 
instruments, \new{from the eras of the OSO missions and SKYLAB, to today after
the routine use of }
adaptive optics for solar physics
around 2000 \citep{Rimmele+others1999,Scharmer+others1999}.   

\new{Finally, we note that
detailed analysis of polarized light in strong ground-based lines 
continues, with diverse goals. Recent works
have studied individual chromospheric multiplets 
with photospheric lines.  As well as H$\alpha$  \citep[see Section 4 of][]{JaumeBestard+others2022},
the 8542 \AA{} line of Ca$^+$ 
continues to attract attention
\nnew{as a diagnostic of chromospheric magnetism}}.
\citet{Gosic+others2021} detected upward extensions of photospheric internetwork fields, and 
\citet{Vissers+others2022}  
tested extrapolated fields
using this line.   \nnew{Chromospheric 
magnetic fields inferred from polarization measurements of 
 helium 10830 \AA{} were combined with 
intensity  data from GREGOR and IRIS.
Interstingly, no correlation of magnetic properties (including electric currents) 
with enhanced heating was found.}

To date, these and most previously published polarimetric
studies used data from 
one chromospheric line or multiplet, often observed
with photospheric lines.  
The work presented here
emphasizes the need to
study 
the optimal lines (Mg$^+$)
simultaneously with multiple other 
 lines, probing 
magnetic and thermal structure 
along each line of sight through the chromosphere.   The combination of Mg$^+$ and
Fe$^+$ polarimetry therefore deserves serious consideration for future experimental efforts
\citep{Judge+others2021}.  

\bigskip
\noindent 
\section*{Data  availability statement}

The  research reported here used IDL-based 
software developed by the first author without documentation.  The ``data'' produced are exploratory in
nature.  Interested readers 
can request outputs from PJ.

\section*{Acknowledgments}

The authors are grateful to Joan Burkepile, 
Rebecca Centeno Elliot, Yuhong Fan, Holly Gilbert, Giuliana de Toma, Anna Malanushenko, and Matthias Rempel for inspiration and 
discussion which have led to
the present work.  This
material is based upon work supported by the National Center for
Atmospheric Research, which is a major facility sponsored by the
National Science Foundation under Cooperative Agreement No. 1852977.

\bibliographystyle{aasjournal}
\bibliography{ms}
\end{document}

%% file: defs.tex
\newcommand{\be}[1]{\begin{equation} \label{eq:#1}}
\newcommand{\ee}{\end{equation}}
\newcommand{\ba}[1]{\begin{eqnarray} \label{eq:#1}}
\newcommand{\ea}{\end{eqnarray}}

\newcommand{\B}{\ifmmode
\mathbf{B}\else {\bf B}\fi}

\newcommand{\solrad}{\ifmmode{R}_{\rm S}\else${R}_{\rm S}$\fi}
\newcommand{\solmas}{\ifmmode{M}_{\rm S}\else${M}_{\rm S}$\fi}

\newcommand{\tintu}{\ifmmode{\rm erg~cm^{-2}~s^{-1}sr^{-1}}\else 
  erg~cm$^{-2}$~s$^{-1}$~sr$^{-1}$\fi}
\newcommand{\fluxu}{\ifmmode{\rm erg~cm^{-2}~s^{-1}}\else 
  erg~cm$^{-2}$~s$^{-1}$\fi}

\newcommand{\wave}{\ifmmode{\lambda} \else$\lambda$\fi}

\newcommand\lta { \mathrel {\hbox to 0pt {\lower 3.7pt \hbox{$\sim$}
      \hss} \raise 1.7pt \hbox{$<$}}}
\newcommand\gta { \mathrel {\hbox to 0pt {\lower 3.7pt \hbox{$\sim$}
      \hss} \raise 1.7pt \hbox{$>$}}}

\newcommand{\intu}{\ifmmode{\rm erg~cm^{-2}~s^{-1}sr^{-1} \AA^{-1}}\else 
 erg~cm$^{-2}$~s$^{-1}$~sr$^{-1}$
 ~\AA$^{-1}$\fi}

\newcommand{\intunu}{\ifmmode{\rm erg~cm^{-2}~s^{-1}sr^{-1} \AA^{-1}}\else 
 erg~cm$^{-2}$~s$^{-1}$~sr$^{-1}$
 ~Hz$^{-1}$\fi}
\newcommand\nnew[1]{#1}
\newcommand\new[1]{#1}

\newcommand{\figstart}{
\begin{figure*}
\includegraphics[width=0.8\linewidth]{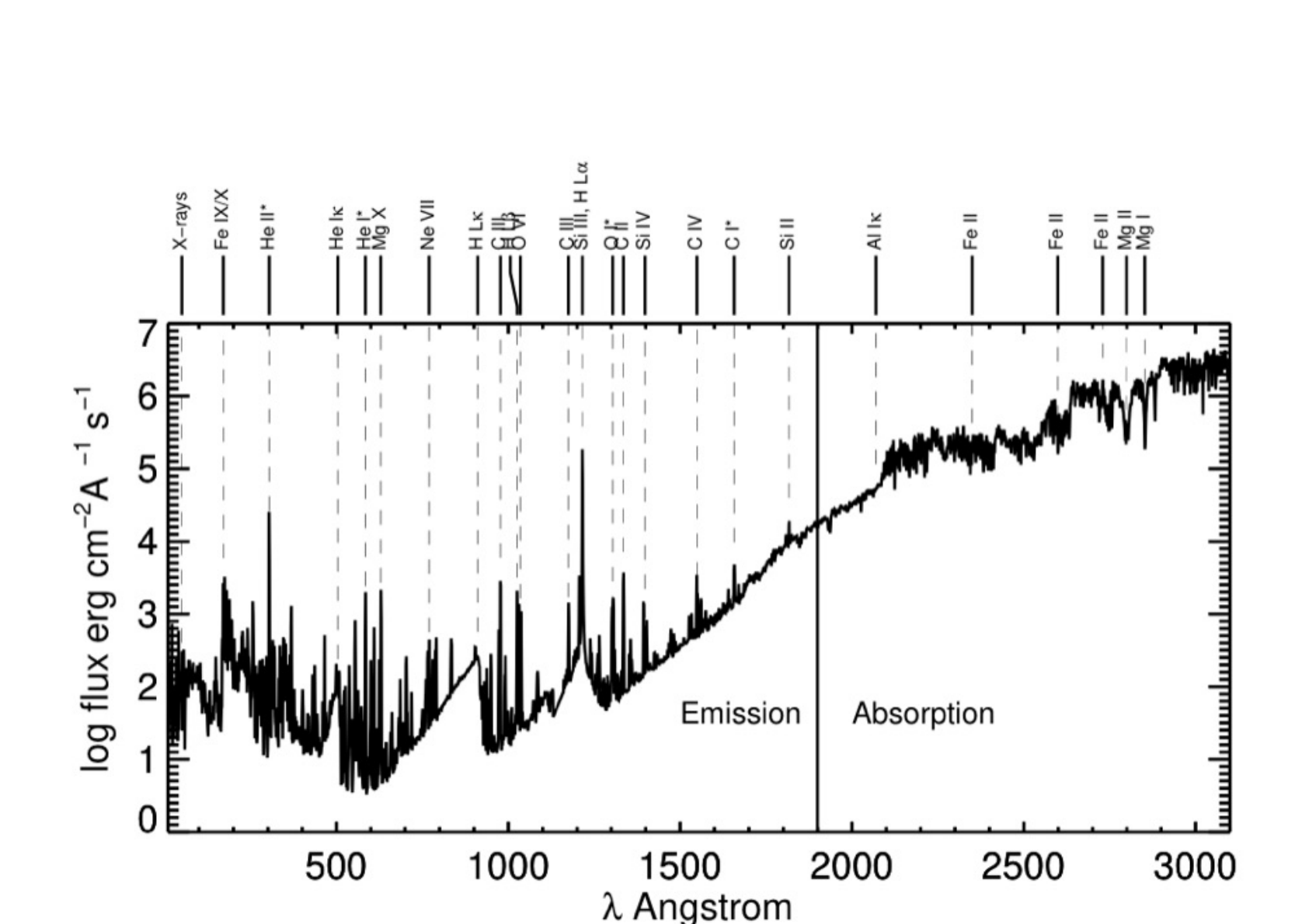}  
\caption{The solar flux spectrum 
as measured on Mar 25-29 2008 with various
instruments, compiled by staff at LASP
as part of the Whole Heliospheric 
Interval Campaign, from \url{http://lasp.colorado.edu/lisird/}.
The spectrum is from soft 
X-rays
to near the atmospheric cut-off,
highlighting emission and absorption features. 
The region has a number of deep absorption lines with emission in their core.
Mg II is such an example, although the emission is not observable due to the modest spectral resolution. The transitions
of H, He I and He II, with $\Delta n=1$ and are labeled with asterisks are compared with $\Delta n=0$ transitions of
much more highly ionized species at similar wavelengths.  } 
\label{fig:start} 
\end{figure*}
}

\newcommand{\figcomp}{
\begin{figure}
\includegraphics[width=\linewidth]{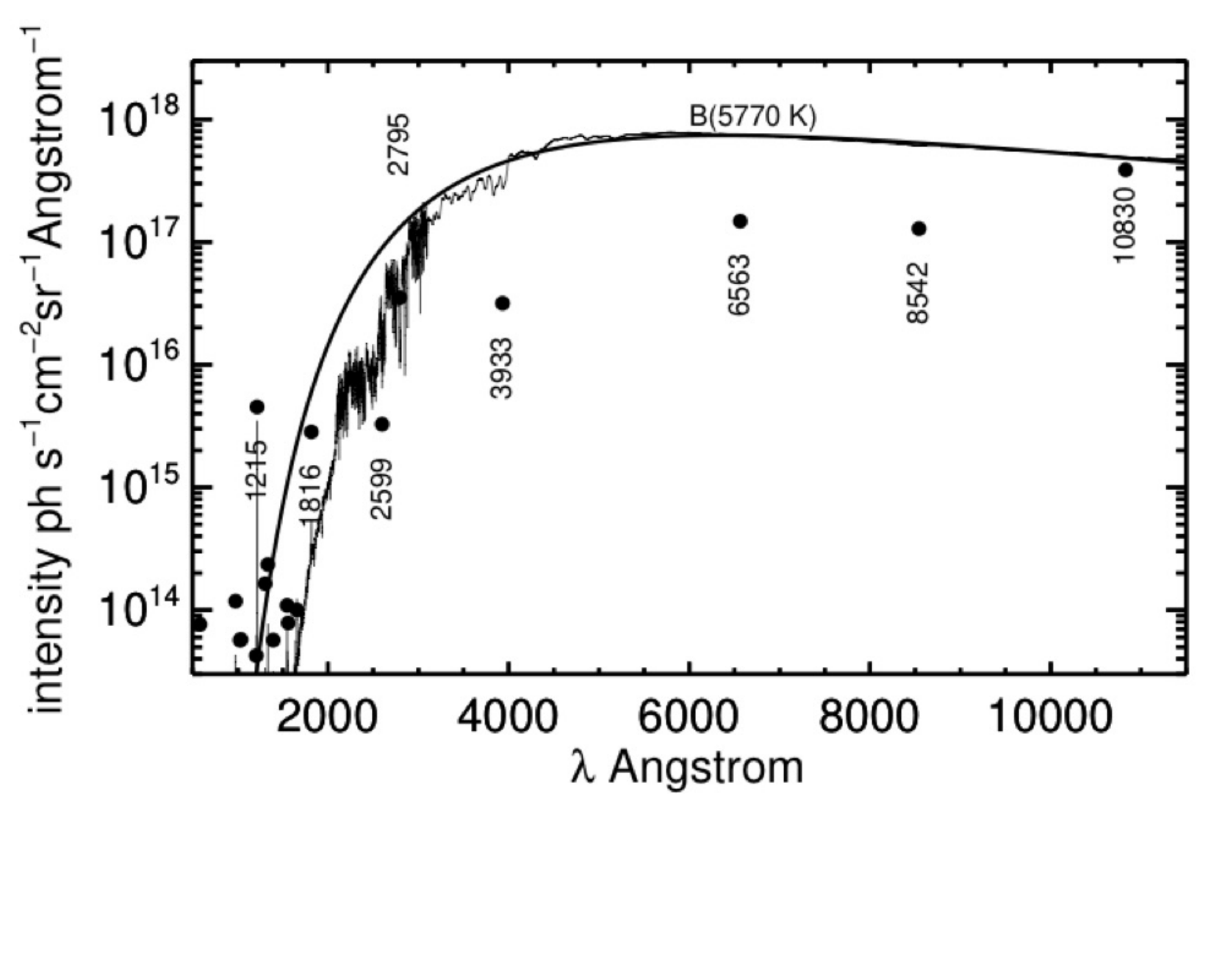}  
\caption{Intensities in the cores of the strongest lines are plotted, \new{together with a black body spectrum at the solar 
effective temperature for reference.  \new{Also shown is 
a modest (typically 2\AA) resolution intensity spectrum from the Whole Heliosphere Interval campaign
(\url{https://lasp.colorado.edu/lisird/data/whi_ref_spectra}), derived as $1/\pi$ of the irradiance spectrum scaled to
the solar surface.}
Line core intensities (filled circles) used here and their sources are listed in Table~\ref{tab:calcs}.}}
\label{fig:comp} 
\end{figure}
}

\newcommand\fillit{\hspace{ 14pt}}

\newcommand\tabstrong{
{ \protect 
  \begin{table*}
      \caption{Strong spectral lines in the  solar spectrum} 
      \begin{tabular}{lrllcccccrl}
      \hline\hline
      Ion & $\lambda$ & Iso & I.P. & I$_{0}$ &
      $-\log \frac{\tau}{\tau_{L\alpha}}$ & $\overline g$&$\overline G$ & $g_L$& $B_H$& Notes\\
          &     \AA{} & 
          &   eV & & &\multicolumn{2}{c}{ Zeeman } &Hanle& G&\\
 		      \hline
H &1215.674&&13.6&7.4(4)&0.3& 1.33 & 1.33& \\
&1215.688&&&&0& 1.17 & 1.33& 1.33 & 53$^a$ \\
 &1025.722& &&1.1(3)&0.8& 1.17& 1.33 &1.33& 14$^a$\\ 
 &1025.723& &&&1.1& 1.33& 1.33 && \\
  &  6563\fillit{} &  & &3.5(5) & $\approx$ 5.5-7.6 &0.83-1.33 &0.69-1.33 &0.67-2.00&1.5-7.7
  & 6 blended lines \\
He &584.334&&24.6&2.6(3)&1.1& 1.00 & 1.00& 1.00 & 200\\
   & 10829.084& &&8(5)& $\approx10.2$ &2.00 & 4.01& \\
   & 10830.243& &&7(5)&$\approx9.8$  &1.75 &2.88&1.50 & 0.87\\
   & 10830.333& &&7(5)&$\approx9.5$ &1.25 &1.53&1.50&0.87\\
He$^+$ &304.780&H&54.4&1.6(3)&2.1&1.33&1.33& \\
& 304.786&&& &1.8&1.17&1.33&1.33 & 850\\
C&1560\fillit{} & &11.3&1.0(3)&4.2-6.1&0.5-2.0 &0.25-3.86&0.50-1.33& 0.74-15 & 6 blended lines\\
C&1657\fillit{} & &&1.2(3)&3.9-4.5&1.5 &2.25& 1.00-1.50 &6.5-19& 6 blended lines\\
C$^{+}$&1334.532&B&24.4&3.5(3)&4.2&0.83 &0.69&0.80&34\\
&1335.663&&&&4.9&1.07&0.72&0.80& 6.8\\
&1335.708&&&&3.9&1.10 &1.21&1.20 & 27\\
C$^{2+}$&977.020&Be&47.9&2.4(3)&3.5&1.0&1.0&1.00& 200 \\
C$^{3+}$&1548.187&Li&64.5&1.4(3)&3.8&1.17&1.33& 1.33&23\\
N$^{4+}$&1238.821&Li&77.5&2.2(2)&5.0& 1.17&1.33& 1.33&29\\
O&1302.168&&13.6&2.5(3)&3.6&1.25 &1.53&2.00&19\\
&1304.858&&&2.5(3)&3.8&1.75 &2.88&2.00&12\\
&1306.029&&&2.5(3)&4.3&2.00 &4.01&2.00&3.8\\
O$^{5+}$&1031.912&Li&138&1.1(3)&4.2& 1.17&1.33&1.33&36\\
Ne$^{7+}$&770.409&Li&239&2.4(2)&5.7& 1.17&1.33&1.33&50\\
Mg$^+$&2795.528&Na&15.0&2.5(5)&3.7& 1.17& 1.33&1.33& 23 &  Mg~II $k$ line\\
Mg$^{9+}$&609.973&Li&367&5.2(1)&6.0& 1.17&1.33&1.33&65 \\
Si$^{+}$&1816.926&Al&16.3&3.0(4)&6.4&1.10&1.21& 1.20& 0.28
\\
Si$^{2+}$&1206.500&Mg&33.5&7.0(2)&3.9&1.00&1.00&1.00& 300\\
Si$^{3+}$&1393.775&Na&45.1&8.1(2)&4.9&1.17&1.33& 1.33&79\\
Ca$^+$&3933.663&K&11.9&2.5(5)&4.7& 1.17&1.33&1.33 &13      & Fraunhofer $K$ line\\
  & 8542.086 & &&2.0(5) &6.1 & 1.10 & 1.21& 1.33& 0.84\\
Fe$^+$&2585-2633&Mn&16.2&2.5(4)&5.0-6.9&1.50-1.87 &2.25-2.75&1.55-1.87&0.2-11& 11 unblended lines\\
\hline
		          \end{tabular} 
		          \break
\protect\flushleft{\small The UV lines listed are those identified in  figure~\ref{fig:start}, with $I_0$  line core intensities in the quiet Sun (7.4(4) $\equiv 7.4 \cdot 10^4$) erg~cm$^{-2}$ s$^{-1}$ \AA$^{-1}$,  optical 
depths relative to H L$\alpha$ 
for a 100 km path length, calculated using  electron and total hydrogen densities of $10^{11}$ cm$^{-3}$
and assuming all of
the element populations are in the listed ionization stage. The
L$\alpha$ optical depth at line center is $10^6$. Electron temperatures of
8000 K were used for 
neutrals and for lines of Si$^+$,
Mg$^+$, Ca$^+$ and Fe$^+$, otherwise 
peak temperatures of ``coronal''
ionization equilibrium were used
\citep{Jordan1969}.  Other
sources:
The Mg$^{9+}$ intensity is from \cite{Vernazza+Reeves1978};
Between 670 and 1611 \AA{}, intensities  are from 
\cite{Curdt+others2001};  for 
multiplets, the highest 
intensity line alone is listed;  intensities for Si$^+$ and Fe$^+$ are taken from 
the Hubble flux spectrum of $\alpha$ Cen A assuming 
no limb brightening
\citep{Ayres2010},
\new{the line center intensity of Si$^+$ 1816.9 \AA{} agrees satisfactorily with estimates derived from photographic solar data 
\citep{Nicolas+others1977}};
the Mg$^+$ $k$ line intensity is from \cite{Staath+Lemaire1995};
  Ca$^+$ K line intensity is 
from \cite{Ayres+others1976}, from $\alpha$ Cen A, 
it is almost identical to 
solar values.  $^a$\cite{Raouafi+others2016}, all other 
Hanle fields $B_H$ are from this work. Lines with no $g_L$ and 
$B_H$ values are  transitions blended with lines of interest, but where the upper level is unpolarizable ($J=0$ or 1/2) and so no
Hanle effect. 
} 
\label{tab:calcs} 
  \end{table*}
}
}